\documentclass{article}
\usepackage{graphicx}
\usepackage{natbib}
\usepackage{xurl}
\usepackage{amsmath}
\usepackage[hidelinks]{hyperref}
\usepackage{float}
\usepackage{caption}
\usepackage{booktabs}
\usepackage{multirow}

\title{How Many Significant LISA?: A multiple testing correction}
\author{Giuseppe Arbia\thanks{Corresponding author. Email: \href{mailto:giuseppe.arbia@unicatt.it}{giuseppe.arbia@unicatt.it}}
  \qquad Francesca Cardile \\
  Vincenzo Nardelli \qquad Niccol\`o Salvini \\[1.5ex]
  \normalsize Universit\`a Cattolica del Sacro Cuore, Milan, Italy}
\hypersetup{
  pdftitle={How Many Significant LISA? A multiple testing correction},
  pdfauthor={Giuseppe Arbia, Francesca Cardile, Vincenzo Nardelli, Niccolo Salvini}}
\date{}

\begin{document}

\maketitle

\begin{abstract}
In spatial data analysis, global measures of spatial correlation provide useful insights into the general spatial patterns, but fail to capture local variations. Local Indicators of Spatial Association (LISA) address this limitation by identifying local clusters and spatial outliers, offering valuable applications in many fields of research such as public health, urban planning and environmental monitoring, to name only a few. However, the application of LISA tools in hypothesis testing is  undermined by  the presence of the multiple testing problem which dramatically increases the risk of committing Type I errors especially when dealing with very large datasets. Standard corrections procedure (like, e. g., Bonferroni and Benjamini-Hochberg procedures) fail to consider the spatial dependence among the various LISA tests and, furthermore, they are overly conservative. This paper proposes an adjustment that, by incorporating the notion of the ``effective sample size'', improves statistical inference balancing false discovery rate control and statistical power.
\end{abstract}

\vspace{1em}

\section{Introduction}
In the era of big data, it is crucial to employ tools that can effectively summarize the geographical features of spatial datasets. The analysis of spatial data often begins with the calculation of global measures of spatial autocorrelation that provide insight into overall spatial patterns. Among these, Moran's I \citep{moran1950}, Geary's C \citep{geary1954} and APLE \citep{LiCalderCressie2007} have long been considered fundamental tools in spatial statistics.

In particular, Moran’s I has been widely applied in various disciplines, including regional economics, public health \citep{waller2004}, environmental monitoring \citep{haining2003}, and criminology \citep{leitner2011}. These global indices are essential to determine whether a spatial process exhibits a general clustering of similar (or dissimilar) values in a given study area. However, global measures often fail to capture local variations in spatial patterns, which are crucial for identifying clusters or anomalies. This limitation has motivated the development and widespread adoption of the so-called Local Indicators of Spatial Association (LISA), introduced by \citet{anselin1995}.

A popular version of the LISA index is represented by the so-called local Moran coefficient (LMC) \citep{anselin1995}. Generally speaking, LISA statistics allow a deeper understanding of spatial data distribution by identifying local clusters (of the high-high or low-low typology) and spatial outliers (of the high-low or low-high typology). 

The applications of LISA methods are vast and span a wide range of disciplines, including regional science, public health, economics, environmental studies, epidemiology, criminology, and urban planning. One of the earliest and most influential applications of LISA tests was provided by \citet{ReyMontouri1999}, who illustrated their usefulness in regional economics by revealing local clusters and spatial outliers in U.S. regional income convergence, thereby showing that convergence dynamics are intrinsically spatial rather than geographically random. Many relevant applications may be found also in public health surveillance \citep{getis1992}. For example, LISA statistics have been extensively used to study the spatial distribution of COVID-19 cases, revealing high-risk cluster patterns that informed targeted intervention strategies \citep{Siljander2022} which proved to be of tremendous importance during the pandemic, when spatial surveillance played a crucial role in monitoring disease transmission. In environmental science, LISA statistics have been employed to analyse air pollution hotspots and water quality variations \citep{lee2001}. In social sciences and urban studies, LISA maps have been used to investigate patterns of income inequality and social deprivation \citep{kneebone2013}. Furthermore, LISA has been utilized in criminology (to identify areas of high crime and evaluate the impact of crime prevention policies \citep{leitner2011}), in
 urban planning \citep{fotheringham2002} and many others.
Given the foregoing examples, it becomes clear that the validity of the statistical inference underlying LISA significance tests is not merely a theoretical issue but a fundamental requirement for evidence-based decision making. Consequently, any shortcomings in the inferential framework may have important practical implications on the conclusions drawn from LISA analyses.

The discussion of these shortcomings is exactly the concern of the present paper. 

As a matter of fact, despite their wide applicability, when used in an inferential context, LISA statistics present significant challenges related to the issue of the "multiple hypothesis testing" \citep{Shaffer1995}. Indeed, when running a large number of significance tests simultaneously (as it happens in the LISA analysis), the risk of Type I errors (false positives) increases dramatically.

The multiple testing problem is well documented in classical statistics and some corrections have been introduced to overcome it \citep{Bonferroni1936,benjamini1995,WestfallYoung1993}. However, the conventional corrections methods are not applicable in the spatial context not only because they assume independence of the tests (which is obviously violated in this case), but also because they are highly conservative, leading to under-rejection of the null hypotheses especially when dealing with a large number of tests \citep{dunn1961}.
Even if several contributions can be found in the literature that tackle the problem of dependence (e. g. \citet{BenjaminiYekutieli2001, Wu2008, Sarkar2002,BlanchardRoquain2009}), the problem of multiple testing in spatially correlated local statistics still remains largely unexplored.

In this paper, we propose an adjustment of the standard  corrections procedures which accounts for spatial dependence and that provides a balance between controlling the false discovery rate and preserving statistical power, thus improving the reliability of LISA-based spatial inferences.

The remainder of the paper is organized as follows. In Section~2, we employ a widely used spatial dataset to illustrate the misleading conclusions that may arise from LISA significance tests when the multiple-testing problem is ignored. In Section 3 we provide  an overview of the multiple testing problem and of the standard correction procedures. In Section 4, we introduce the notion of effective sample size in spatial samples and we show how we can leverage on it in order to adjust the standard procedures thus tackling the multiple testing problem with spatially distributed datasets. Section 5 discusses the results of a Monte Carlo experiment aiming at showing the inferential properties of our proposed method, while  Section 6 is devoted to discussing a case study. Finally, Section 7 concludes.

\section{A motivating example to start: The Colombian malaria dataset}
To motivate the problem addressed in this paper, let us consider a popular example of the use of LISA that led to relevant substantial conclusions.

In particular, we refer to the well-known dataset on malaria incidence in Colombia available from the \textit{GeoDa Data Lab} \citep{GeoDaMalaria2005}. The Colombian malaria dataset has become a standard benchmark in the GeoDa ecosystem for illustrating local spatial autocorrelation methods and LISA cluster maps. Its widespread use as a teaching and methodological dataset, together with its clear spatial clustering pattern, makes it an ideal benchmark for investigating the inferential properties of LISA significance tests. The dataset consist of malaria incidence rates observed in 1998 for 1,066 Colombian municipalities. Figure 1 reports the LISA map highlighting the (5\%) significant municipalities.
\begin{figure}[htbp]
\centering
\includegraphics[width=0.65\textwidth]{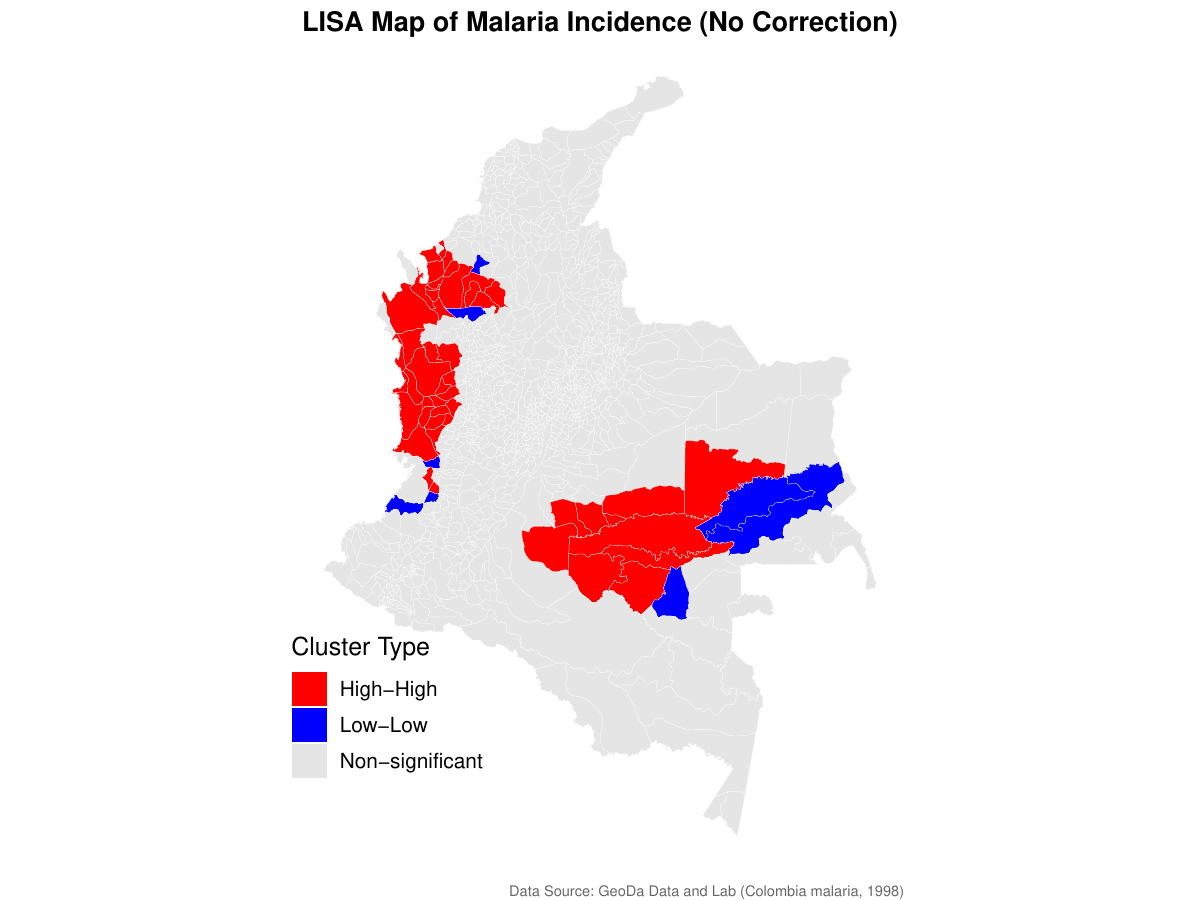}
\caption{LISA map of malaria incidence across the 1,066 Colombian municipalities (1998). Standard local Moran tests identify 47 High--High clusters and 8 Low--Low clusters before any correction for multiple testing.}
\label{fig:malaria_lisa}
\end{figure}
A standard LISA analysis identifies 55 clusters (47 High--High  and 8 Low--Low), suggesting the presence of substantial local spatial clustering. In particular, the LISA analysis identifies 47 hot spot with clear indications for health policy targeting of resources.
However, these results immediately raise an important inferential question. Since a separate significance test is performed for each municipality, the analysis involves 1,066 simultaneous hypothesis tests. Even if all null hypotheses were true, a non-negligible number of municipalities would be expected to appear significant purely by chance. For example, if we consider a 0.05 level of significance, 53 municipalities (that is most of the 55 significant LISA!) can potentially be wrongly included in the hot/cold spot critical location, leading to inefficient targeting of resources.  
Consequently, the observed clusters cannot be interpreted uncritically without accounting for the multiple-testing problem.

This example illustrates the central question addressed in the remainder of the paper: how can we distinguish genuine local clusters and outliers from false discoveries?

\section{The problem of multiple testing}
\label{sec:htmt}

Hypothesis testing is a fundamental concept in statistical inference, aiming to determine whether observed patterns in data provide sufficient evidence to reject a null hypothesis in favour of an alternative hypothesis. In the context of spatial analysis, hypothesis testing often involves assessing spatial autocorrelation significance, detecting clusters, or identifying spatial outliers. 

The challenge of multiple testing arises when several hypotheses are tested simultaneously, increasing the likelihood of Type I errors (false positives). Traditional methods for controlling multiple comparisons include the Bonferroni correction \citep{Bonferroni1936} and the Benjamini-Hochberg procedure \citep{benjamini1995}. 

In particular, the Bonferroni Correction (BC) is a conservative approach that adjusts the significance level by dividing it by the number of tests:
  \[
  \alpha_{\text{Bonferroni}} = \frac{\alpha}{m},
  \]
where $\alpha$ is the designated significance level and \( m \) is the total number of hypotheses tested. This method ensures a stringent control over Type I errors, but may lead to increased Type II errors. 

In contrast with the Bonferroni correction, the Benjamini-Hochberg (BH) procedure is a false discovery rate (FDR) control method that ranks p-values \( p_{(1)}, p_{(2)}, ..., p_{(m)} \) in ascending order and selects a threshold \( k \) such that:
  \[
  p_{(k)} \leq \frac{k}{m} \alpha.
  \]
The BH procedure is less conservative than the BC and provides a better balance between detecting true effects and controlling false discoveries.

Both the BC and the BH corrections are based on the assumption that the tests are independent. The problem of multiple testing becomes considerably more challenging when the hypothesis of independence is violated, as is typically the case for spatial, temporal, and spatio-temporal data. Ignoring the dependence structure may lead to substantial distortions in the actual error rates, resulting either in an inflation of false discoveries or in overly conservative inference. Consequently, considerable effort has been devoted to extending multiple-testing procedures to dependent data. A major breakthrough was provided by \citet{BenjaminiYekutieli2001}, who proved that the Benjamini--Hochberg procedure controls the False Discovery Rate (FDR) under the Positive Regression Dependency on a Subset (PRDS) condition and proposed a conservative adjustment that guarantees FDR control under arbitrary dependence. Subsequently, \citep{Wu2008} introduced a conditional dependence framework that explicitly models dependence among hypotheses, deriving asymptotic properties of the False Discovery Proportion and demonstrating that dependence can be exploited rather than ignored, particularly in applications involving spatially structured hypotheses. Further theoretical developments include the asymptotic analysis of weakly dependent tests by \citet{Farcomeni2007}, who showed that standard FDR procedures remain valid under broad classes of weak dependence, as well as the decision-theoretic framework of \citet{SunCai2009}, which incorporates dependence through hidden Markov models and achieves substantial gains in statistical power while maintaining FDR control. Other important contributions include the generalized stepwise procedures under positive dependence proposed by \citet{Sarkar2002}, the unified dependency-control framework of \citet{BlanchardRoquain2009}, and resampling-based approaches that explicitly exploit the dependence structure among test statistics \citep{WestfallYoung1993}.

Despite these many important advances, most existing methods have been developed for dependence structures arising in genomics, signal processing, or time series, whereas the problem of multiple testing in spatially correlated local statistics, such as LISA statistics, remains comparatively less explored. 
For this reason, we devote the next section to introduce an operational procedure to correct the LISA hypothesis testing for the problem of multiple testing.

\section{The "Effective Sample Size" correction}

Spatial data are intrinsically characterized by dependence among neighboring observations, implying that the information provided by each spatial unit is not entirely independent of that contained in its geographical neighborhood. As a consequence, in the presence of positive spatial dependence, the effective amount of statistical information conveyed by a spatial sample is generally smaller than its nominal size. To account for this phenomenon, \citet{cressie1993} introduced the concept of the \emph{equivalent number of independent observations}, also referred to as the \emph{effective sample size} (henceforth ESS). This quantity represents the number of independent observations that would provide approximately the same inferential precision as the original spatially correlated sample.  Following Cressie, the effective sample size can be expressed as a function of both the nominal sample size $n$ and the degree of spatial autocorrelation, measured by the APLE coefficient $\rho$ \citep{LiCalderCressie2007}. 
Formally this can be expressed as follows:

\begin{equation}\label{eq:ess}
    \text{ESS} = \frac{n}{1 + 2 \left( \frac{\rho}{1 - \rho} \right) \left( 1 - \frac{1}{n} \right) - 2 \left( \frac{\rho}{(1 - \rho)^2} \right) \left( \frac{1 - \rho^{(n-1)}}{n} \right)}.    
\end{equation}

 In the words of Cressie; ``If $n = 10$ and $\rho = 0.26$, then  ESS = 6.2; that is, 6 independent observations achieve approximately the same precision as 10 correlated observations with $\rho = 0.26$'' 
\citep{cressie1993}.
As a consequence, when the level of positive spatial autocorrelation increases, the effective sample size decreases monotonically, reflecting the growing redundancy of the information contained in neighboring observations. In the limit, as $\rho$ approaches unity, the effective sample size tends to zero, indicating that the sample contains progressively less independent information. This limiting behavior closely resembles the inferential consequences of spatial unit root processes discussed by \citet{MuellerWatson2024}, where strong spatial persistence gives rise to spurious statistical inference and requires dedicated testing procedures.

Furthermore, notice that, in the asymptotic case, the effective sample size reduces approximately to:

    \[
    \text{ESS} = n\,\frac{1-\rho}{1+\rho}
    \]

clearly showing that, even in very large samples, spatial dependence may substantially reduce the amount of independent information available for statistical inference. 

The notion of the effective sample size is derived by considering the spatial autocorrelation
structure within the data: instead of treating each observation as independent,
ESS estimates the number of truly independent units within the dataset by
incorporating a correction factor based on a spatial autocorrelation measure. This correction ensures that the degrees of freedom are
adjusted to reflect the true number of independent observations rather than the
total number of data points. This observation has important practical implications from the multiple testing procedures point of view. Indeed, classical corrections such as Bonferroni and Benjamini--Hochberg implicitly assume that all tests are independent and therefore use the nominal sample size when adjusting significance thresholds. In the presence of spatial dependence, however, these procedures become unnecessarily conservative. Motivated by this consideration, our proposal consists of replacing the \textit{nominal} sample size with the \textit{effective} sample size in the computation of both Bonferroni and Benjamini--Hochberg corrections, thereby explicitly accounting for the information redundancy induced by spatial autocorrelation while preserving the rationale underlying the original procedures.
It is important to delimit precisely what the proposed adjustment does and does not address. Our correction operates on the \emph{multiplicity} of the local tests: it takes the marginal $p$-value attached to each location as given, and modifies the threshold against which the whole family of $p$-values is compared. A logically distinct question concerns the validity of those marginal $p$-values themselves. The reference distribution normally used for the local Moran statistic, whether obtained analytically under the randomisation assumption or by conditional permutation, corresponds to a null hypothesis of complete spatial randomness, in which the observed values are exchangeable across locations. When the underlying process displays global spatial autocorrelation, that null is violated everywhere, and the local statistic is referred to a distribution that does not describe the process generating the data. This problem was identified and addressed by \citet{OrdGetis2001}, who propose a local statistic that remains valid in the presence of global spatial autocorrelation. The two issues are complementary rather than alternative: the ESS adjustment developed here corrects the multiplicity of a family of local tests, while the approach of \citet{OrdGetis2001} corrects the calibration of each individual test. Combining the two -- applying a multiplicity correction based on the effective sample size to locally calibrated $p$-values -- appears to us a natural and promising direction, and we return to it in the concluding section.

The effects of this modification on the BC and on the BH procedures will be tested in the next two sections, respectively through a Monte Carlo exercise and through the re-examination of the malaria dataset presented in Section 2.

\section{A Monte Carlo experiment}

\subsection{Design of the experiment}
\label{sec:montecarlo}

To assess the finite-sample performance of the proposed adjusted multiple-testing corrections, we conduct a Monte Carlo simulation study in which the true locations of local spatial association are known by construction. The simulation design is intended to reproduce a controlled spatial setting while preserving the marginal distribution of the simulated variable. This allows us to isolate the ability of alternative testing procedures to correctly identify local clusters and spatial outliers without confounding the results with changes in the underlying distribution of the data.

We consider a regular square lattice consisting of $40\times40$ spatial units, yielding a total sample size of $n=1{,}600$. Spatial dependence is defined through a first-order Queen contiguity matrix, according to which two spatial units are considered neighbors whenever they share either a common boundary or a vertex. Let $W$ denote the corresponding spatial weights matrix, which is row-standardized throughout the analysis.

For each Monte Carlo replication, the baseline data-generating process is a spatial lag model (SLM) \citep{Arbia2024},

\begin{equation}
\mathbf{y} = \rho W\mathbf{y} + \boldsymbol{\varepsilon},
\label{eq:slm_mc}
\end{equation}

where

\begin{equation}
\boldsymbol{\varepsilon} \sim N(\mathbf{0},\sigma^2 I_n).
\end{equation}

Equivalently,

\begin{equation}
\mathbf{y}
=
(I_n-\rho W)^{-1}\boldsymbol{\varepsilon}.
\label{eq:slm_reduced}
\end{equation}

The spatial autoregressive parameter $\rho$ is allowed to vary over the range between $0$ and $0.95$, thereby giving us the possibility to evaluate the performance of the corrections under progressively stronger levels of positive spatial dependence. The case $\rho = 0$ provides a useful reference corresponding to spatially independent observations, since the reduced form~\eqref{eq:slm_reduced} then collapses to $\mathbf{y}=\boldsymbol{\varepsilon}$, while values of $\rho$ approaching $1$ generate substantial spatial persistence. The upper limit $\rho=1$ is not attainable: with a row-standardised weights matrix the largest eigenvalue of $W$ equals one, so $(I_n-W)$ is singular and the reduced form does not exist. The value $\rho=0.95$ therefore represents the near-unit-root end of the admissible parameter space, which is precisely the regime discussed by \citet{MuellerWatson2024}. This dimension of the experiment is particularly important because stronger spatial dependence reduces the effective amount of independent information contained in the lattice and therefore directly affects the multiplicity problem underlying local tests of spatial association.

After generating the baseline spatial process, we introduce a controlled set of local spatial patterns using a \emph{marginal-preserving permutation} procedure. The key feature of this procedure is that no new observations are generated and no values are modified in each run of the simulation. Instead, the values already present in the simulated sample are simply selectively reassigned across spatial locations. Consequently, the empirical marginal distribution of $\mathbf{y}$ is exactly preserved, while its local spatial configuration is altered in a controlled manner.
A related approach was employed by \citet{CaldasDeCastroSinger2006} to investigate multiple-testing corrections for local statistics of spatial association under spatial dependence.

In our exercise, starting from the baseline dataset,  we then construct four types of local spatial association corresponding to the conventional quadrants of the Moran scatterplot, namely: (i) high--high (HH) clusters, (ii) low--low (LL) clusters, (iii) high--low (HL) and (iv) low--high (LH) spatial outliers. To generate HH clusters, observations belonging to the upper tail of the empirical distribution are relocated to a set of neighboring spatial units. Similarly, LL clusters are generated by reallocating observations from the lower tail to contiguous locations. The resulting configurations therefore contain compact areas in which unusually high or unusually low values are spatially concentrated.

Spatial outliers are generated according to the opposite principle. Selected extreme observations are placed in locations surrounded predominantly by observations from the opposite tail of the distribution. In particular, HL configurations consist of relatively high observations located within predominantly low-valued neighborhoods, whereas LH configurations consist of relatively low observations surrounded by high-valued neighbors. The locations used for these spatial outliers are selected randomly in each simulation run, subject to the requirement that the intended local configuration is obtained.
(See Figure 2).

\begin{figure}[ht!]
        \centering
        \includegraphics[width=1\textwidth]{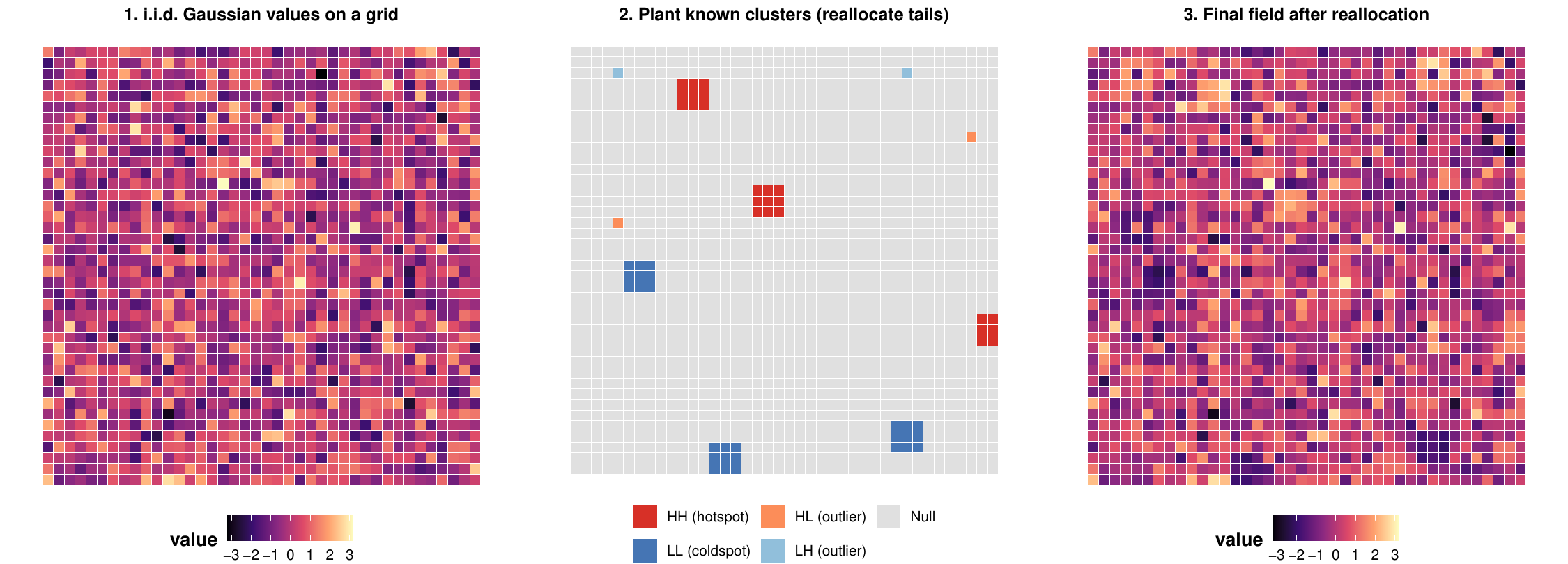}
        \caption{The planting mechanism, illustrated on the $40\times40$ lattice used in the experiment and, for clarity, for the case $\rho=0$. Left pane: the baseline field. Middle pane: the ten planted configurations, whose locations constitute the ground truth. Right pane: the final field, which carries the planted structure with exactly the same marginal distribution as the baseline, since only the spatial allocation of the values has changed.}
        \label{fig:planting}
    \end{figure}

An important feature of this experimental design is that the manipulation affects only the \emph{spatial allocation} of the observations. 
If $\mathbf{y}^{(0)}$ denotes the original realization generated according to equation~\eqref{eq:slm_reduced}, the manipulated realization can be expressed as

\begin{equation}
\mathbf{y}^{(*)}=P\mathbf{y}^{(0)},
\label{eq:permutation}
\end{equation}

where $P$ is an $n\times n$ permutation matrix constructed so as to generate the desired local configurations. Since $P$ only changes the ordering of the observations,

\begin{equation}
\left\{y^{(*)}_1,\ldots,y^{(*)}_n\right\}
=
\left\{y^{(0)}_1,\ldots,y^{(0)}_n\right\},
\end{equation}

as multisets. Thus, all statistics that depend exclusively on the marginal empirical distribution---including the sample mean, variance, quantiles, and higher-order empirical moments---remain unchanged. What changes is exclusively the spatial arrangement of the observations and, consequently, their local spatial association.

This construction provides a particularly convenient framework for evaluating Local Indicators of Spatial Association (LISA). Because the units into which the tail observations are deliberately relocated are recorded, the ground-truth status of each spatial unit is known exactly. Each location can therefore be classified as belonging to an imposed HH or LL cluster, an imposed HL or LH spatial outlier configuration, or to the background component of the lattice. The simulation can consequently be interpreted as a spatial signal-detection experiment in which the true set of local alternatives is known.

For each simulated dataset, local Moran statistics are computed for all $n=1{,}600$ spatial units and the corresponding local significance tests are performed. We then compare the unadjusted LISA procedures with the conventional multiple-testing corrections (BC and BH) and with our proposed adjustments, which consider the ESS in place of the nominal sample size. In particular, the simulation design makes it possible to distinguish between two dimensions of performance. The first concerns the ability to detect the deliberately introduced local patterns, while the second concerns the ability to avoid declaring significance at locations that do not belong to the imposed ground-truth configurations.

Let $\mathcal{S}$ denote the set of locations belonging to the imposed spatial patterns and $\widehat{\mathcal{S}}$ the set of locations detected as significant by a given testing procedure. The number of true and false discoveries can then be defined respectively as:

\begin{equation}
TP = |\widehat{\mathcal{S}}\cap\mathcal{S}|,
\qquad
FP = |\widehat{\mathcal{S}}\cap\mathcal{S}^{c}|.
\end{equation}

Similarly, false negatives and true negatives are given by

\begin{equation}
FN = |\widehat{\mathcal{S}}^{c}\cap\mathcal{S}|,
\qquad
TN = |\widehat{\mathcal{S}}^{c}\cap\mathcal{S}^{c}|.
\end{equation}

These quantities allow us to assess both error control and detection power. 
Consequently, in order to compare the performance of the alternative multiple-testing correction procedures, 
we consider two complementary evaluation metrics, namely: \emph{Power}
and the \emph{False Discovery Rate} (FDR). These measures are constructed from
the quantities introduced above, namely the number of true positives ($TP$), 
false positives ($FP$), false negatives ($FN$), and true negatives ($TN$). 
Since the locations of the artificially imposed spatial patterns are known 
by construction, these quantities can be computed exactly in each Monte 
Carlo replication.

The first measure is the \emph{Power}, defined as

\begin{equation}
    \mathrm{Power} =
    \frac{TP}{TP+FN}.
    \label{eq:power}
\end{equation}

Power measures the ability of a testing procedure to detect the local spatial 
patterns deliberately introduced into the simulated data. A value close to 
one therefore indicates that the procedure is able to recover most of the 
imposed local spatial signals, whereas a low value indicates that a substantial 
fraction of the true patterns remains undetected.

The second measure is the \emph{False Discovery Rate} (FDR). For each Monte 
Carlo replication, we first define the False Discovery Proportion (FDP) as

\begin{equation}
    FDP =
    \begin{cases}
        \dfrac{FP}{TP+FP} & \text{if } TP+FP>0,\\[1.5ex]
        0 & \text{if } TP+FP=0,
    \end{cases}
    \label{eq:fdp}
\end{equation}

where the second case ensures that the proportion is well defined 
when no hypothesis is rejected. The FDR is then defined as the expected value 
of the FDP,

\begin{equation}
    FDR = E(FDP).
    \label{eq:fdr}
\end{equation}

In the Monte Carlo experiment, this expectation is estimated by averaging 
the FDP across the replications. FDR therefore measures the expected 
proportion of false discoveries among all locations declared significant. 
Unlike Power, which focuses on missed signals, FDR focuses on the contamination 
of the set of detected locations by false positives. Its complement,
$1-FDP$, measures the reliability of the set of discoveries, so that a single
quantity summarises both aspects of the trade-off and no separate measure of
precision needs to be reported.

Taken together, the two measures provide a comprehensive assessment of the
performance of the alternative correction procedures. Power captures their
ability to recover the imposed spatial signals, while FDR quantifies the
expected proportion of false discoveries among the locations they declare
significant. The comparison is
particularly relevant in the present setting because a less conservative
multiple-testing correction may increase Power while simultaneously
increasing FDR. The two metrics therefore allow us to
evaluate explicitly the trade-off between signal detection and error control
generated by the proposed ESS adjustment.

The entire experiment is replicated independently for each value of $\rho$ and for each considered spatial configuration. Reported performance measures are obtained by averaging the corresponding quantities across Monte Carlo replications. By varying $\rho$ while keeping the lattice structure, marginal distribution, and ground-truth mechanism fixed, the experiment provides direct evidence on how increasing spatial dependence affects the performance of alternative multiple-testing corrections for local spatial association.

Overall, the proposed design separates three components that are often intertwined in empirical applications: the marginal distribution of the variable, the strength of global spatial dependence, and the presence of specific local spatial configurations. In particular, the marginal-preserving permutation strategy ensures that differences in detection performance cannot be attributed to changes in the distribution of the observations. They arise instead from the interaction between the imposed local spatial signal, the underlying degree of spatial dependence, and the multiple-testing correction under consideration.

\subsection{Simulation Results}
\label{subsec}

Figure 3 summarizes the performance of the alternative procedures under increasing levels of background spatial autocorrelation, from exact spatial independence ($\rho=0$) to the near-unit-root case ($\rho=0.95$). A clear trade-off emerges between detection power and error control, and this trade-off becomes progressively more pronounced as the spatial autoregressive parameter $\rho$ increases. Table~\ref{tab:mc_results} reports the corresponding figures.

\begin{figure}[ht!]
        \centering
        \includegraphics[width=1\textwidth]{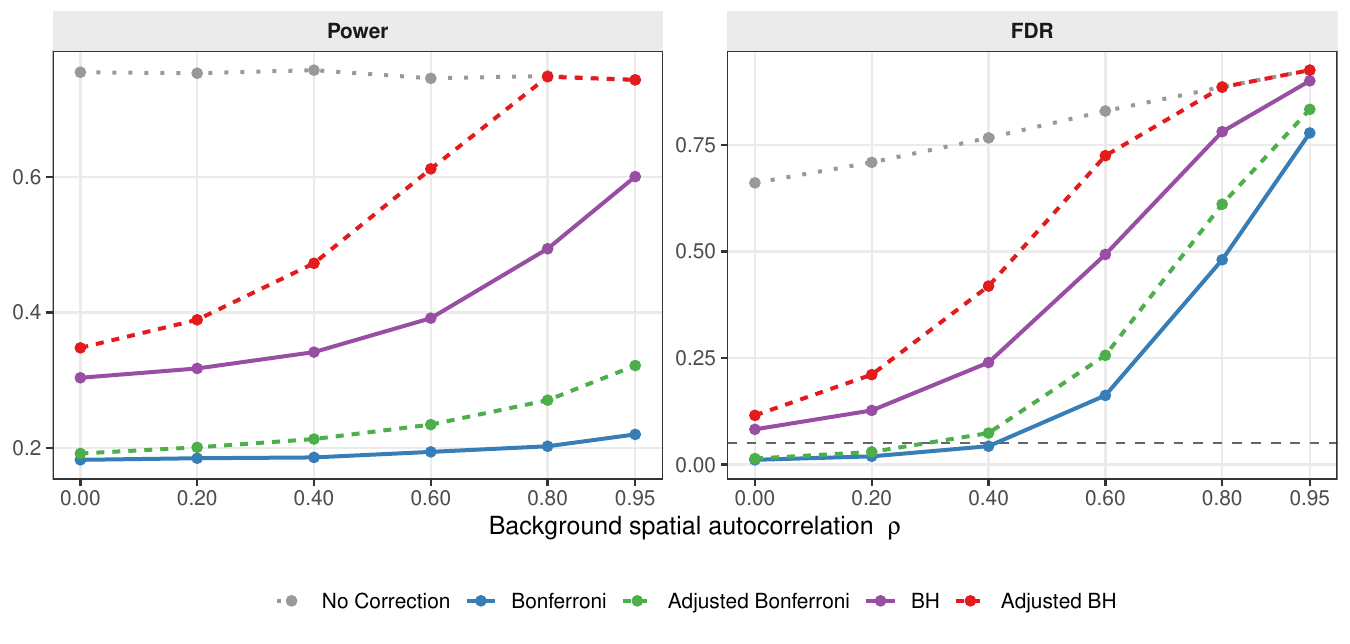}
\caption{Detection power and realized FDR of the four correction procedures, together with the uncorrected LISA benchmark, as a function of the background spatial autocorrelation $\rho$. The dashed line in the right-hand panel marks the nominal $5\%$ level.}
        \label{fig:metrics}

    \end{figure}

\begin{table}[ht!]
\centering
\small
\begin{tabular}{lrrrrrr}
\hline
Method & $\rho=0$ & $\rho=0.2$ & $\rho=0.4$ & $\rho=0.6$ & $\rho=0.8$ & $\rho=0.95$ \\
\hline
\multicolumn{7}{l}{\emph{Power}} \\
No Correction & 0.755 & 0.753 & 0.758 & 0.746 & 0.749 & 0.743 \\
Bonferroni & 0.182 & 0.184 & 0.186 & 0.194 & 0.202 & 0.220 \\
Adjusted Bonferroni & 0.191 & 0.201 & 0.213 & 0.234 & 0.270 & 0.321 \\
BH & 0.303 & 0.317 & 0.341 & 0.391 & 0.494 & 0.600 \\
Adjusted BH & 0.347 & 0.389 & 0.472 & 0.612 & 0.748 & 0.743 \\
\hline
\multicolumn{7}{l}{\emph{False Discovery Rate}} \\
No Correction & 0.661 & 0.709 & 0.766 & 0.829 & 0.886 & 0.925 \\
Bonferroni & 0.011 & 0.019 & 0.043 & 0.162 & 0.480 & 0.778 \\
Adjusted Bonferroni & 0.014 & 0.030 & 0.074 & 0.256 & 0.610 & 0.833 \\
BH & 0.082 & 0.127 & 0.239 & 0.493 & 0.781 & 0.900 \\
Adjusted BH & 0.115 & 0.211 & 0.418 & 0.724 & 0.885 & 0.925 \\
\hline
$ESS/n$ & 0.675 & 0.540 & 0.410 & 0.287 & 0.174 & 0.110 \\
$\widehat{\rho}$ (APLE) & 0.196 & 0.300 & 0.420 & 0.555 & 0.705 & 0.803 \\
\hline
\end{tabular}
\caption{Monte Carlo results as a function of the background spatial autocorrelation $\rho$: detection power and realized false discovery rate of the five procedures, averaged over $500$ replications ($40\times40$ lattice, $n=1{,}600$, $\alpha=0.05$). The last two rows report the average estimated effective sample size relative to the nominal one and the average APLE estimate.}
\label{tab:mc_results}
\end{table}

\subsubsection*{The case of spatial independence}

The case $\rho=0$ is the natural calibration benchmark, since in the absence of background spatial dependence the ESS adjustment has, in principle, nothing to exploit. The results confirm that in this situation the adjusted procedures behave almost exactly like their conventional counterparts. Adjusted Bonferroni attains a power of $0.191$ against $0.182$ for conventional Bonferroni, with realized FDR of $0.014$ against $0.011$; adjusted BH attains $0.347$ against $0.303$, with FDR of $0.115$ against $0.083$. The Bonferroni-based procedures remain well below the nominal $5\%$ threshold, as expected from a family-wise criterion applied to an FDR benchmark.

This is a reassuring property of the proposed adjustment: where spatial dependence is absent, replacing the nominal sample size with the effective one leaves the inferential conclusions essentially unchanged, so that nothing is lost by adopting the adjusted procedures as a default.

It should be stressed, however, that even at $\rho=0$ the estimated effective sample size does not coincide with the nominal one. The APLE coefficient estimated on the simulated fields equals $0.196$ on average, corresponding to $ESS/n=0.675$. The reason is that the effective sample size is estimated from the data, and the data contain the planted configurations themselves: a lattice carrying ten local clusters exhibits a non-negligible amount of spatial autocorrelation even when the underlying process is spatially independent. The ESS correction is therefore mildly active also in this limiting case.

\subsubsection*{Detection power}

Conventional Bonferroni is by far the most conservative procedure. Its power remains approximately constant at around $0.19$ over most of the parameter space, rising only marginally to $0.220$ at $\rho=0.95$. This flatness is precisely what one would expect from a correction that divides the significance level by the nominal number of tests irrespective of the dependence structure: the threshold does not react to the information content of the sample.

The conventional BH procedure performs better, with power increasing from $0.303$ at $\rho=0$ to $0.600$ at $\rho=0.95$. The proposed adjustments yield further gains throughout. Adjusted Bonferroni improves upon its conventional counterpart by a margin that widens monotonically with $\rho$, from $0.191$ to $0.321$, an increase of roughly $46\%$ at the highest level of dependence considered. Adjusted BH exhibits the largest gains, rising from $0.347$ at $\rho=0$ to $0.748$ at $\rho=0.8$.

The mechanism is transparent. As $\rho$ grows, the estimated effective sample size falls steadily, from $ESS/n=0.675$ at $\rho=0$ to $0.174$ at $\rho=0.8$ and $0.110$ at $\rho=0.95$. Since the adjusted thresholds are computed on the effective rather than on the nominal number of tests, they become progressively less stringent exactly when the conventional corrections become progressively more conservative relative to the amount of independent information actually available.

\subsubsection*{False discovery rate}

These gains in power are accompanied by a deterioration in error control, and the realized FDR increases monotonically in $\rho$ for every procedure. Under weak background dependence all the corrected procedures remain reasonably close to the nominal $5\%$ benchmark, with the Bonferroni-based procedures comfortably below it up to $\rho=0.4$ ($0.043$ and $0.074$ respectively) and the BH-based ones exceeding it but remaining moderate ($0.239$ and $0.418$).

As $\rho$ increases, however, the FDR rises sharply for all procedures. At $\rho=0.8$ it reaches $0.480$ for Bonferroni, $0.611$ for adjusted Bonferroni, $0.781$ for BH and $0.885$ for adjusted BH. The deterioration is systematically steeper for the adjusted procedures than for their conventional counterparts, and steepest of all for adjusted BH, which is also the procedure achieving the largest gains in power. The ordering of the four procedures is the same in both panels of Figure 3 and simply reversed: what is gained in detection is paid for in the reliability of the resulting set of discoveries.

\subsubsection*{The near-unit-root case}

The behaviour at $\rho=0.95$ deserves separate comment, because it reveals the natural boundary of the proposed approach. At this level of dependence the adjusted BH procedure becomes numerically indistinguishable from the uncorrected LISA analysis: both attain a power of $0.743$ and an FDR of $0.925$.

The reason is mechanical rather than accidental. The adjusted BH threshold for the $k$-th ordered p-value is $\min\left(k\,\alpha/ESS,\ \alpha\right)$, and at $\rho=0.95$ the estimated effective sample size is approximately $176$ out of $1{,}600$ nominal observations. For every rank beyond the $176$-th the threshold therefore saturates at $\alpha$, and the step-up procedure reduces to testing each location at the nominal significance level without any multiplicity correction at all.

Far from being an anomaly, this is the logical endpoint of the argument developed in Section~4. If the effective number of independent observations collapses towards a small figure, then the effective multiplicity of the testing problem collapses with it, and there is progressively less multiplicity left to correct for. The same phenomenon does not occur for adjusted Bonferroni, whose threshold $\alpha/ESS$ remains a genuine correction ($2.8\times10^{-4}$ at $\rho=0.95$) and which consequently still contains the FDR at $0.833$ against $0.925$ for the uncorrected analysis. This asymmetry suggests that under very strong spatial persistence the family-wise version of the adjustment is the more prudent choice.

\subsubsection*{Interpretation}

Taken together, these results reveal an important feature of local multiple testing under spatial dependence. Adjusting the corrections increases the probability of detecting the imposed local signals, but the gain in power does not translate into an unambiguous improvement in inferential performance. As background spatial autocorrelation increases, extreme observations outside the deliberately constructed clusters become increasingly likely to exhibit locally coherent spatial configurations and are consequently classified as significant. The distinction between the imposed ``ground-truth'' signal and spatial association generated endogenously by the background SLM process therefore becomes progressively less sharp. This mechanism explains the simultaneous increase in power and false discoveries observed in Figure 3.

It should be stressed, in this respect, that a location classified as a false positive because it lies outside the artificially imposed ground-truth patterns may nevertheless exhibit genuine local spatial association generated endogenously by the background SLM process. The point applies with particular force to the ring of locations immediately surrounding a planted cluster: these units are adjacent to a genuine concentration of extreme values and therefore display real local association by construction, yet the ground truth records them as null. This is one reason why the realized FDR of the BH-based procedures exceeds the nominal level already at $\rho=0$. Consequently, and especially for large values of $\rho$, the empirical FDR reported here should be interpreted relative to the imposed ground truth rather than as a measure of purely spurious statistical significance.

With this qualification in mind, the practical reading of Figure 3 is the following. Up to moderate levels of background spatial dependence, say $\rho\leq0.4$, the ESS adjustment delivers a systematic increase in detection power at a cost in error control that remains modest, and it is in this range that we would recommend its use. Beyond that threshold the adjusted procedures, and adjusted BH in particular, become increasingly permissive, until in the near-unit-root case they cease to correct for multiplicity altogether. The effective sample size therefore provides not only an operational device for relaxing excessively conservative thresholds, but also an explicit indication of when local inference on spatially dependent data ceases to be informative.

\section{The Colombian malaria dataset revisited}
\label{sec:casestudy}

We now return to the Colombian malaria dataset discussed in Section~2 to illustrate the empirical implications of the proposed adjustments. As discussed previously, the dataset contains malaria incidence rates observed in 1998 for $n=1,066$ Colombian municipalities. The conventional LISA analysis, conducted at the 0.05 significance level without correcting for multiple testing, identifies 55 significant local clusters, consisting of 47 High--High (HH) and 8 Low--Low (LL) municipalities. These results provide a useful benchmark against which the alternative multiple-testing procedures can be compared.

Before applying the proposed corrections, we assess the degree of spatial dependence characterizing the data. Global Moran's $I$ is equal to $0.3189$ and is highly significant ($p=0.001$), providing strong evidence of positive global spatial autocorrelation. More importantly for the adjustment proposed in this paper, the APLE spatial autocorrelation coefficient is estimated as $\widehat{\rho}=0.4911$. Applying the correction introduced in Equation 1 provides an evaluation of the effective sample size equal to 364.5 which is substantially smaller than the nominal sample size of 1,066 municipalities. Thus, according to the ESS criterion, the information contained in the spatial sample is approximately equivalent to that provided by only 365 independent observations. This large reduction reflects the substantial redundancy of information induced by positive spatial dependence and provides the empirical motivation for replacing the nominal number of tests with the effective number of independent spatial units in the adjusted multiple-testing procedures.

Table~\ref{tab:lisa_counts} compares the results obtained from the uncorrected LISA analysis with those produced by the conventional Bonferroni (BC) and Benjamini--Hochberg (BH) corrections and by their spatially adjusted counterparts.

\begin{table}[ht!]
    \centering
    \resizebox{\textwidth}{!}{
    \begin{tabular}{lcccccc}
        \toprule
        & \multicolumn{3}{c}{Counts} & \multicolumn{3}{c}{Percentage} \\
        \cmidrule(lr){2-4}\cmidrule(lr){5-7}
        Method & High-High & Low-Low & Non-signif. & High-High & Low-Low & Non-signif. \\
        \midrule
        No Correction           & 47 & 8 & 1011 & 4.41\% & 0.75\% & 94.84\% \\
        Bonferroni (BC)         & 26 & 3 & 1037 & 2.44\% & 0.28\% & 97.28\% \\
        Adjusted BC             & 27 & 4 & 1035 & 2.53\% & 0.38\% & 97.09\% \\
        Benjamini-Hochberg (BH) & 34 & 7 & 1025 & 3.19\% & 0.66\% & 96.15\% \\
        Adjusted BH             & 35 & 8 & 1023 & 3.28\% & 0.75\% & 95.97\% \\
        \bottomrule
    \end{tabular}}
    \caption{Comparison of the uncorrected LISA analysis with the conventional
    Bonferroni (BC) and Benjamini--Hochberg (BH) corrections and with their
    ESS-adjusted counterparts, for the $1{,}066$ Colombian municipalities.}
    \label{tab:lisa_counts}
\end{table}

As expected, the conventional Bonferroni correction produces the most conservative outcome. The number of significant municipalities falls from 55 to 29, of which 26 are classified as HH and only 3 as LL. In other words, almost one half of the local associations identified by the uncorrected analysis are removed once strict FWER control is imposed. Although such conservativeness is intrinsic to the Bonferroni principle, in a spatial setting it may be further amplified by treating all 1,066 local tests as if they conveyed independent information.

Replacing the nominal number of tests with the effective sample size partially compensates for this effect. The adjusted Bonferroni procedure identifies 31 significant municipalities, comprising 27 HH and 4 LL locations. The numerical difference with respect to conventional Bonferroni is relatively modest, but it occurs in the expected direction: accounting for spatial dependence relaxes the excessively stringent threshold generated by using the nominal sample size and recovers additional local associations.

The difference is more apparent when considering the Benjamini--Hochberg procedure. Conventional BH identifies 41 significant municipalities, consisting of 34 HH and 7 LL locations. When the BH correction is adjusted using the effective sample size, the number increases to 43, with 35 HH and all 8 LL clusters identified by the original uncorrected analysis. Thus, among the procedures considered, adjusted BH provides the closest approximation to the spatial structure revealed by the baseline LISA map while still explicitly addressing the multiple-testing problem.

The geographical distribution of the significant municipalities is also informative (see Figure 4). 

\begin{figure}[H]
    \centering
    \includegraphics[width=\textwidth]{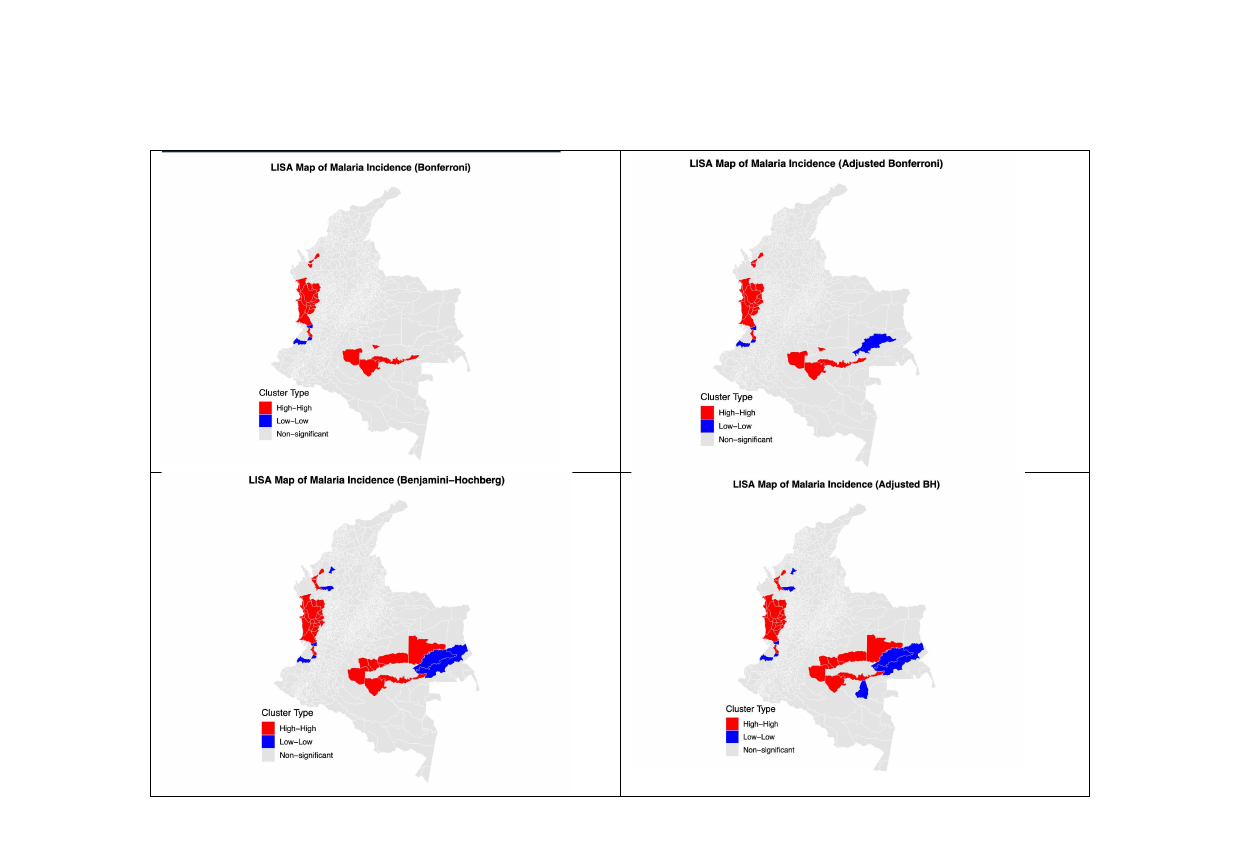}
    \caption{Comparison of the geographical distribution of the four correction methods.}
    \label{fig:maps}
\end{figure}

In fact, the comparison between the maps obtained under the four correction procedures and the one reported in Figure 1 shows that the principal areas of spatial concentration identified by the original LISA analysis are broadly preserved. Indeed the Bonferroni Correction retains only the strongest components of these clusters, whereas BH produces a spatial configuration closer to the uncorrected map. The ESS-adjusted procedures recover additional municipalities primarily within, or in close proximity to, these pre-existing spatial concentrations rather than generating a radically different geographical pattern. This suggests that the adjustment operates mainly at the margins of the detected clusters, where significance is most sensitive to the choice of the multiple-testing threshold.

These results illustrate an important practical implication of accounting for spatial dependence in multiple testing. The conventional corrections implicitly treat the $1,066$ local tests as if they represented the same amount of independent information as $1,066$ unrelated statistical tests. The estimated ESS of only $364.5$ indicates that this assumption is particularly restrictive in the present application. Once the redundancy induced by spatial autocorrelation is explicitly incorporated into the correction mechanism, the significance threshold becomes less conservative and some of the local associations discarded by the conventional procedures are recovered.

At the same time, the adjusted procedures should not be interpreted as an attempt to reproduce the uncorrected LISA map. The latter provides a descriptive benchmark but does not address the multiplicity problem that motivated this study. Rather, the purpose of the ESS adjustment is to retain the logic of multiple-testing correction while recognizing that the effective multiplicity of a spatially dependent system may be substantially smaller than its nominal number of locations. From this perspective, the empirical application illustrates the compromise underlying the proposed approach: conventional Bonferroni provides the strongest protection against false positives at the cost of a substantial reduction in detected local associations; BH is less conservative; and the ESS-adjusted versions recover part of the potentially informative spatial signal by explicitly incorporating the dependence structure of the data.

For the Colombian malaria data, this distinction is substantively relevant. As emphasized in Section~2, the identification of HH municipalities may have direct implications for the geographical targeting of public-health interventions. An excessively liberal procedure risks directing attention and resources toward false hotspots, whereas an excessively conservative procedure may fail to identify municipalities belonging to genuine high-incidence clusters. The adjusted procedures are designed precisely to mitigate this trade-off by incorporating the effective amount of independent spatial information into the multiple-testing correction.

Overall, the empirical evidence is consistent with the rationale underlying the proposed methodology. Positive spatial autocorrelation substantially reduces the effective sample size, and explicitly accounting for this reduction moderates the conservativeness of conventional multiple-testing corrections. In the present application, the adjustment produces a modest recovery of significant locations under Bonferroni and a somewhat larger set of retained local associations under BH, while preserving the main geographical structure of the clusters identified in the original LISA analysis.

\section{Conclusions and future research challenges}
\label{sec:conclusions}

Local Indicators of Spatial Association have become a standard tool for identifying geographically concentrated phenomena in regional science and spatial data analysis. Their inferential use, however, raises a fundamental multiple-testing problem: when a local test is performed at every spatial unit, the probability of obtaining false discoveries can increase substantially. Conventional multiple-testing corrections address this problem but do not explicitly account for the spatial dependence linking neighboring observations and tests. As a consequence, particularly in large and strongly autocorrelated spatial datasets, they may impose corrections that are excessively conservative.

This paper proposes a simple adjustment based on the concept of the effective sample size (ESS). The central idea is to replace the nominal number of spatial units entering conventional Bonferroni and Benjamini--Hochberg corrections with an effective number of independent observations that reflects the redundancy of information induced by spatial dependence. The resulting procedure preserves the basic logic and operational simplicity of standard multiple-testing corrections while explicitly introducing the strength of spatial dependence into the determination of the significance threshold.

The Monte Carlo evidence highlights both the potential and the limitations of this approach. The ESS-adjusted procedures systematically increase the ability to detect the deliberately imposed local spatial patterns relative to their conventional counterparts. The gain is particularly pronounced for the adjusted Benjamini--Hochberg procedure and becomes larger as positive spatial dependence increases. This result is consistent with the intuition underlying the ESS correction: as spatial autocorrelation becomes stronger, the nominal number of observations increasingly overstates the amount of independent information contained in the sample, making conventional corrections progressively more restrictive.

At the same time, the simulation results reveal a clear trade-off between detection power and error control. The additional power obtained through the adjustment is accompanied by an increasing empirical FDR as background spatial dependence becomes stronger. This finding calls for a cautious interpretation. In our simulation design, the ground truth refers to the local configurations deliberately introduced through the marginal-preserving permutation mechanism. Under strong spatial dependence, however, locations outside these imposed clusters may themselves develop genuine local spatial association as a consequence of the underlying spatial process. Thus, a location classified as a false positive relative to the imposed ground truth is not necessarily a purely spurious statistical discovery. This distinction becomes increasingly important as the spatial autoregressive parameter approaches the upper boundary of its admissible range.

The empirical application to malaria incidence across Colombian municipalities confirms the practical relevance of the proposed adjustment. Strong positive spatial dependence reduces the nominal sample of $1,066$ municipalities to an estimated effective sample size of approximately $365$. Conventional Bonferroni and BH corrections substantially reduce the number of significant LISA relative to the uncorrected analysis, while their ESS-adjusted counterparts recover some additional local associations. Importantly, these additional discoveries tend to occur within, or close to, the main spatial concentrations already identified by the alternative procedures, rather than generating entirely new geographical patterns. The application therefore illustrates the practical compromise motivating our approach: accounting for multiplicity while avoiding an unnecessarily severe loss of potentially informative local spatial signals.

Several limitations of the present analysis also point toward directions for further research. A first important issue concerns \emph{negative spatial autocorrelation}. The present paper focuses primarily on positive spatial dependence, which is the most common setting considered in empirical applications and corresponds to the intuitive case in which spatial correlation reduces the amount of independent information in the sample. Negative spatial autocorrelation raises a substantially different problem. As discussed by \citet{GriffithArbia2010} and \citet{Griffith2019}, negatively autocorrelated spatial processes are characterized by neighboring observations tending to be dissimilar and have received considerably less attention than their positively autocorrelated counterparts. In such circumstances, the interpretation of ESS as a simple measure of information loss due to spatial redundancy becomes less immediate and, depending on the dependence structure, the effective amount of information may behave very differently from the positive-correlation case. Extending the proposed multiple-testing adjustment to negative spatial dependence therefore requires a careful reconsideration of both the definition and the inferential interpretation of spatial effective sample size.

A second issue arises at the opposite extreme of the parameter space, when positive spatial dependence becomes very strong. In the ESS formulation considered here, increasing $\rho$ progressively reduces the effective number of independent observations, with the ESS approaching its lower bound as $\rho$ approaches unity. This behavior establishes an interesting connection with the literature on \emph{spatial unit roots}. In particular, \citet{MuellerWatson2024} show that highly persistent spatial processes may generate inferential phenomena analogous to those associated with unit roots in time series, including spurious statistical relationships and non-standard asymptotic behavior. From this perspective, the collapse of the ESS under near-unit-root spatial dependence provides an information-based interpretation of the same phenomenon: although the nominal number of spatial observations may remain large, the amount of independent information available for inference may become extremely small. The Monte Carlo experiment makes this visible: at $\rho=0.95$ the estimated effective sample size falls to about $176$ out of $1{,}600$ nominal locations, and the adjusted Benjamini--Hochberg threshold consequently saturates at the nominal significance level, so that the procedure ceases to correct for multiplicity altogether and coincides with the uncorrected LISA analysis. The effective sample size thus signals not only how much the conventional thresholds should be relaxed, but also the point beyond which local inference on strongly persistent spatial data ceases to be informative. Investigating more formally the relationship between effective sample size, local multiple testing, and spatial unit-root asymptotics represents an important direction for future theoretical research.

A third limitation is the assumption that the strength of spatial dependence can be adequately summarized by a single global parameter. Real regional systems are frequently characterized by substantial spatial heterogeneity, and the same dataset may contain sub-regions displaying very different degrees---and potentially different signs---of spatial autocorrelation. A natural extension of the present framework would therefore allow the study area to be partitioned into $G$ spatial regimes, each characterized by a regime-specific spatial dependence parameter $\rho_g$ and a corresponding effective sample size $ESS_g$, for $g=1,\ldots,G$. Multiple-testing corrections could then be locally calibrated using the effective amount of independent information available within each spatial regime rather than relying on a single global ESS. Such an extension would be particularly attractive for regional applications in which different geographical areas are governed by distinct spatial mechanisms and would connect the proposed methodology with the broader literature on spatial heterogeneity and spatial regimes \citep{AnselinAmaral2024}.

Relatedly, future work should consider situations in which positive and negative spatial dependence coexist within the same study region. Spatial autocorrelation mixtures have been documented in the literature and may be concealed by a global coefficient that averages heterogeneous local dependence structures. In this setting, a single global ESS could mask important differences in the actual information content of different parts of the spatial system. Developing local or regime-specific measures of effective sample size could therefore provide a more flexible foundation for multiple-testing correction in heterogeneous spatial environments.

A further direction concerns the interaction between multiplicity and the calibration of the individual local test. As noted in Section~4, the multiplicity correction proposed here operates on marginal $p$-values that are themselves obtained under a null hypothesis of complete spatial randomness, an assumption that becomes untenable precisely when global spatial dependence is strong. \citet{OrdGetis2001} provide a local statistic designed for that situation, and coupling their calibration with an effective-sample-size multiplicity correction would address the two sources of distortion simultaneously. We regard this as the most promising extension of the present work.

The same distinction arises outside the LISA framework. Geographically weighted regression estimates one set of coefficients per location and therefore generates a multiple-testing problem of exactly the same nature, with the additional complication that neighbouring local regressions share observations by construction. The issue has been analysed by \citet{SilvaFotheringham2016}, and applications such as \citet{Mollalo2020} routinely interpret maps of locally significant coefficients. Whether an effective-sample-size argument can be transferred to that setting is an open question.

Finally, the proposed framework can be extended naturally from purely spatial LISA to \emph{spatio-temporal LISA} \citep{TaoChenThill2023}.
Indeed, in longitudinal regional datasets, local tests are dependent not only across neighboring geographical units but also across successive time periods. The relevant multiplicity problem therefore operates simultaneously along the spatial and temporal dimensions. A spatio-temporal effective sample size, incorporating both spatial and serial dependence and their interaction, could provide the basis for adjusted multiple-testing procedures for detecting persistent hotspots, emerging clusters, and space--time outliers. Such an extension would be especially relevant for applications involving regional economic dynamics, epidemiological surveillance, environmental monitoring, and other settings in which local spatial patterns evolve over time.

More generally, the results of this paper suggest that the number of statistical tests and the amount of independent information should not be regarded as synonymous in spatial multiple-testing problems. Spatial dependence changes the information content of a geographical sample, and this feature should be reflected in the inferential procedures applied to local statistics. The ESS approach developed here provides a simple first step in this direction. Rather than offering a definitive correction applicable to every form of spatial dependence, it provides a framework for explicitly linking multiple-testing adjustments to the dependence structure of the spatial process. Extending this principle to negative dependence, near-unit-root processes, spatially heterogeneous regimes, and spatio-temporal settings constitutes a promising agenda for further future research.

\section*{Conflict of Interest}
The authors declare no conflict of interest.

\section*{Data Availability Statement}
The data that support the findings of this study are openly available in the
GeoDa Data and Lab repository at
\url{https://geodacenter.github.io/data-and-lab/colomb_malaria/}. The code used
to reproduce the Monte Carlo experiment is available from the corresponding
author upon reasonable request.

\bibliographystyle{apalike}
\bibliography{references}

\begin{thebibliography}{}

\bibitem[Anselin, 1995]{anselin1995}
Anselin, L. (1995).
\newblock Local indicators of spatial association – {LISA}.
\newblock {\em Geographical Analysis}, 27:93--115.

\bibitem[Anselin and Amaral, 2024]{AnselinAmaral2024}
Anselin, L. and Amaral, P.~V. (2024).
\newblock Endogenous spatial regimes.
\newblock {\em Journal of Geographical Systems}, 26(2):209--234.

\bibitem[Arbia, 2024]{Arbia2024}
Arbia, G. (2024).
\newblock {\em A Primer for Spatial Econometrics: With Applications in R}.
\newblock Palgrave Macmillan, Cham, 2 edition.

\bibitem[Benjamini and Hochberg, 1995]{benjamini1995}
Benjamini, Y. and Hochberg, Y. (1995).
\newblock Controlling the false discovery rate: A practical and powerful
  approach to multiple testing.
\newblock {\em Journal of the Royal Statistical Society: Series B},
  57:289--300.

\bibitem[Benjamini and Yekutieli, 2001]{BenjaminiYekutieli2001}
Benjamini, Y. and Yekutieli, D. (2001).
\newblock The control of the false discovery rate in multiple testing under
  dependency.
\newblock {\em The Annals of Statistics}, 29(4):1165--1188.

\bibitem[Blanchard and Roquain, 2009]{BlanchardRoquain2009}
Blanchard, G. and Roquain, E. (2009).
\newblock Adaptive false discovery rate control under independence and
  dependence.
\newblock {\em Journal of Machine Learning Research}, 10:2837--2871.

\bibitem[Bonferroni, 1936]{Bonferroni1936}
Bonferroni, C. (1936).
\newblock Teoria statistica delle classi e calcolo delle probabilità.
\newblock {\em Pubblicazioni del R Istituto Superiore di Scienze Economiche e
  Commerciali di Firenze}, 8:3--62.

\bibitem[Caldas~de Castro and Singer, 2006]{CaldasDeCastroSinger2006}
Caldas~de Castro, M. and Singer, B.~H. (2006).
\newblock Controlling the false discovery rate: A new application to account
  for multiple and dependent tests in local statistics of spatial association.
\newblock {\em Geographical Analysis}, 38(2):180--208.

\bibitem[Cressie, 1993]{cressie1993}
Cressie, N. A.~C. (1993).
\newblock {\em Statistics for Spatial Data}.
\newblock John Wiley \& Sons, New York, revised edition.

\bibitem[da~Silva and Fotheringham, 2016]{SilvaFotheringham2016}
da~Silva, A.~R. and Fotheringham, A.~S. (2016).
\newblock The multiple testing issue in geographically weighted regression.
\newblock {\em Geographical Analysis}, 48(3):233--247.

\bibitem[Dunn, 1961]{dunn1961}
Dunn, O. (1961).
\newblock Multiple comparisons among means.
\newblock {\em Journal of the American Statistical Association}, 56:52--64.

\bibitem[Farcomeni, 2007]{Farcomeni2007}
Farcomeni, A. (2007).
\newblock Some results on the control of the false discovery rate under
  dependence.
\newblock {\em Scandinavian Journal of Statistics}, 34(2):275--297.

\bibitem[Fotheringham et~al., 2002]{fotheringham2002}
Fotheringham, A.~S., Brunsdon, C., and Charlton, M. (2002).
\newblock {\em Geographically Weighted Regression: The Analysis of Spatially
  Varying Relationships}.
\newblock John Wiley \& Sons.

\bibitem[Geary, 1954]{geary1954}
Geary, R.~C. (1954).
\newblock The contiguity ratio and statistical mapping.
\newblock {\em The Incorporated Statistician}, 5:115--146.

\bibitem[{GeoDa Center}, 2005]{GeoDaMalaria2005}
{GeoDa Center} (2005).
\newblock Malaria in {Colombia} (1998).
\newblock \url{https://geodacenter.github.io/data-and-lab/colomb_malaria/}.
\newblock GeoDa Data and Lab. Dataset prepared by Luis Galvis. Accessed August
  2026.

\bibitem[Getis and Ord, 1992]{getis1992}
Getis, A. and Ord, J.~K. (1992).
\newblock The analysis of spatial association by use of distance statistics.
\newblock {\em Geographical Analysis}, 24:189--206.

\bibitem[Griffith, 2019]{Griffith2019}
Griffith, D.~A. (2019).
\newblock Negative spatial autocorrelation: One of the most neglected concepts
  in spatial statistics.
\newblock {\em Stats}, 2(3):388--415.

\bibitem[Griffith and Arbia, 2010]{GriffithArbia2010}
Griffith, D.~A. and Arbia, G. (2010).
\newblock Detecting negative spatial autocorrelation in georeferenced random
  variables.
\newblock {\em International Journal of Geographical Information Science},
  24(3):417--437.

\bibitem[Haining, 2003]{haining2003}
Haining, R. (2003).
\newblock {\em Spatial Data Analysis: Theory and Practice}.
\newblock Cambridge University Press.

\bibitem[Kneebone and Berube, 2013]{kneebone2013}
Kneebone, E. and Berube, A. (2013).
\newblock {\em Confronting Suburban Poverty in America}.
\newblock Brookings Institution Press.

\bibitem[Lee and Wong, 2001]{lee2001}
Lee, J. and Wong, D. (2001).
\newblock Statistical analysis with {ArcView GIS}.
\newblock {\em Journal of Environmental Management}, 62:1--11.

\bibitem[Leitner and Helbich, 2011]{leitner2011}
Leitner, M. and Helbich, M. (2011).
\newblock The impact of hurricanes on crime: A spatio-temporal analysis in the
  city of {Houston, TX}.
\newblock {\em Cartography and Geographic Information Science}, 38:144--156.

\bibitem[Li et~al., 2007]{LiCalderCressie2007}
Li, H., Calder, C.~A., and Cressie, N. (2007).
\newblock Beyond {Moran}'s {I}: Testing for spatial dependence based on the
  spatial autoregressive model.
\newblock {\em Geographical Analysis}, 39(4):357--375.

\bibitem[Mollalo et~al., 2020]{Mollalo2020}
Mollalo, A., Vahedi, B., and Rivera, K.~M. (2020).
\newblock {GIS}-based spatial modeling of {COVID}-19 incidence rate in the
  continental {United States}.
\newblock {\em Science of the Total Environment}, 728:138884.

\bibitem[Moran, 1950]{moran1950}
Moran, P. A.~P. (1950).
\newblock Notes on continuous stochastic phenomena.
\newblock {\em Biometrika}, 37(1/2):17--23.

\bibitem[M{\"u}ller and Watson, 2024]{MuellerWatson2024}
M{\"u}ller, U.~K. and Watson, M.~W. (2024).
\newblock Spatial unit roots and spurious regression.
\newblock {\em Econometrica}, 92(5):1661--1695.

\bibitem[Ord and Getis, 2001]{OrdGetis2001}
Ord, J.~K. and Getis, A. (2001).
\newblock Testing for local spatial autocorrelation in the presence of global
  autocorrelation.
\newblock {\em Journal of Regional Science}, 41(3):411--432.

\bibitem[Rey and Montouri, 1999]{ReyMontouri1999}
Rey, S.~J. and Montouri, B.~D. (1999).
\newblock {US} regional income convergence: A spatial econometric perspective.
\newblock {\em Regional Studies}, 33(2):143--156.

\bibitem[Sarkar, 2002]{Sarkar2002}
Sarkar, S.~K. (2002).
\newblock Some results on false discovery rate in stepwise multiple testing
  procedures under positive dependence.
\newblock {\em Journal of Statistical Planning and Inference}, 102:67--82.

\bibitem[Shaffer, 1995]{Shaffer1995}
Shaffer, J.~P. (1995).
\newblock Multiple hypothesis testing.
\newblock {\em Annual Review of Psychology}, 46:561--584.

\bibitem[Siljander et~al., 2022]{Siljander2022}
Siljander, M., Uusitalo, R., Pellikka, P., Isosomppi, S., and Vapalahti, O.
  (2022).
\newblock Spatiotemporal clustering patterns and sociodemographic determinants
  of {COVID}-19 ({SARS-CoV-2}) infections in {Helsinki, Finland}.
\newblock {\em Spatial and Spatio-temporal Epidemiology}, 41:100493.

\bibitem[Sun and Cai, 2009]{SunCai2009}
Sun, W. and Cai, T. (2009).
\newblock Large-scale multiple testing under dependence.
\newblock {\em Journal of the Royal Statistical Society: Series B},
  71(2):393--424.

\bibitem[Tao et~al., 2023]{TaoChenThill2023}
Tao, R., Chen, Y., and Thill, J.-C. (2023).
\newblock A space-time flow {LISA} approach for panel flow data.
\newblock {\em Computers, Environment and Urban Systems}, 106:102042.

\bibitem[Waller and Gotway, 2004]{waller2004}
Waller, L.~A. and Gotway, C.~A. (2004).
\newblock {\em Applied Spatial Statistics for Public Health Data}.
\newblock John Wiley \& Sons.

\bibitem[Westfall and Young, 1993]{WestfallYoung1993}
Westfall, P.~H. and Young, S.~S. (1993).
\newblock {\em Resampling-Based Multiple Testing: Examples and Methods for
  p-Value Adjustment}.
\newblock John Wiley \& Sons, New York.

\bibitem[Wu, 2008]{Wu2008}
Wu, W.~B. (2008).
\newblock On false discovery control under dependence.
\newblock {\em The Annals of Statistics}, 36(1):364--380.

\end{thebibliography}

\end{document}